\def\aj{AJ}%
\def\apj{ApJ}%
\def\apjl{ApJ}%
\def\aap{A\&A}%
\def\aapr{A\&A~Rev.}%
\def\aaps{A\&AS}%
\def\mnras{MNRAS}%
\def\pasp{PASP}%
\def\nat{Nature}%
\def\kms    {\ifmmode{{\rm km~s}^{-1}}\else{km~s$^{-1}$}\fi}

\def\ccm    {cm$^{-3}$}

\def\Mspy   {\ifmmode {M_{\odot} {\rm yr}^{-1}} \else $M_{\odot}$~yr$^{-1}$\fi}
\def\Mdot   {\ifmmode {\dot M} \else $\dot M$\fi}
\def\mhd    {\ifmmode {n_{{\rm H}_2}} \else $n_{{\rm H}_2}$\fi}
\def\mhcd   {\ifmmode {N_{{\rm H}_2}} \else $N_{{\rm H}_2}$\fi}

\def\El      {\ifmmode{E_{\ell}}\else{$E_{\ell}$}\fi}
\def\beam    {\ifmmode{\theta_{\rm B}}\else{$\theta_{\rm B}$}\fi}
\def\Jyb   {\ifmmode {{\rm Jy~beam}^{-1}} \else{Jy~beam$^{-1}$}\fi}
\def\mjyb   {\ifmmode {{\rm mJy~beam}^{-1}} \else{mJy~beam$^{-1}$}\fi}
\def\mujyb   {\ifmmode {\mu{\rm Jy~beam}^{-1}} \else{$\mu$Jy~beam$^{-1}$}\fi}
\def\Trot   {\ifmmode{T_{\rm rot}}\else$T_{\rm rot}$\fi}

\def\Teff   {\ifmmode{T_{\rm eff}}\else$T_{\rm eff}$\fi}

\def\ITRS   {\ifmmode{\smallint {\rm T}_{R}^{*}dv}\else{$\smallint
{\rm T}_{R}^{*}dv$}\fi}
\def\ITRS   {\ifmmode{\smallint {\rm T}_{R}^{*}dv}\else{$\smallint
{\rm T}_{R}^{*}dv$}\fi}
\def\ITAS   {\ifmmode{\smallint {\rm T}_{A}^{*}dv}\else{$\smallint
{\rm T}_{A}^{*}dv$}\fi}

\def\meth   {CH$_3$OH}
\def\ra	{$\rightarrow$~}

\newbox\grsign      \setbox\grsign=\hbox{$>$}
\newdimen\grdimen   \grdimen=\ht\grsign
\newbox\simgreatbox \setbox\simgreatbox=\hbox{\raise.5ex\hbox{$>$}\llap
                        {\lower.5ex\hbox{$\sim$}}}\ht1=\grdimen\dp1=0pt
\newbox\simlessbox  \setbox\simlessbox =\hbox{\raise.5ex\hbox{$<$}\llap
                        {\lower.5ex\hbox{$\sim$}}}\ht2=\grdimen\dp2=0pt

\def\simless {\mathrel{\copy\simlessbox }}

\documentclass{iaus}
\usepackage{graphicx}

\title[IAU Symposium 287 Summary] %% give here short title %%
{IAU (Maser) Symposium 287 Summary}

\author[Karl M. Menten]   %% give here short author list %%
{Karl M. Menten}

\affiliation{Max-Planck-Institut f\"ur Radioastronomie,\\
Auf dem H\" ugel 69,
53121 Bonn, Germany \\email: {\tt kmenten@mpifr.de}}

\pubyear{2012}
\volume{287}  %% insert here IAU Symposium No.
\pagerange{1--12}
\jname{IAU Symp. 287: Cosmic Masers -- from OH to H$_0$ }
\editors{R. Booth, E. Humphreys, \&\ W. Vlemmings, eds.}
\begin{document}

\maketitle

%\begin{abstract}
%\keywords{Keyword1, keyword2, keyword3, etc.}
%%% add here a maximum of 10 keywords, to be taken form the file <Keywords.txt>
%\end{abstract}

\firstsection % if your document starts with a section,
\section{Introduction}
Almost exactly twenty years ago, the first of a series of conferences dedicated to cosmic masers 
took place in Arlington, Virginia in the USA (March 9--11, 1992) [\cite{1}]. Two more followed, each 
on a 
different 
continent, in Angra dos Reis/Mangaratiba, near Rio de Janeiro, Brasil 
(March 5--10, 2001)  [\cite{2}] and 
in Alice Springs, Australia (March 12--16, 2007)  [\cite{3}].  As at all others meetings,  a large part of 
the 
international maser community convened from January 29 to February 3, 2012 in splendid 
Stellenbosch, 
South Africa, to discuss the state of the art of the field.

Here I'm trying to summarize the many contributions made in 70 oral presentations and by 45 
posters. I 
take the freedom of expanding my discussion of certain topics to provide a larger context for which 
I also 
give some references. Adhering, roughly, to the order of themes as defined by the 
sessions at the 
meeting,
I'm trying to be comprehensive, but I apologize beforehand for any undue omissions and 
misrepresentations. 
 
\section{\label{theory}Maser Theory and Supernova Remnants}
Like many other a maser meeting, this one started with a theory session. An excursion to the 
thermodynamical basics of maser action [\ra S{\scriptsize TRELNITSKI}] was followed by a 
demonstration 
of how complex things can get by the description of class II methanol (\meth) maser excitation via 
mid-infrared pumping, cycling through the molecule's torsionally excited levels [\ra S{\scriptsize 
OBOLEV} channeled by G{\scriptsize RAY}].

New ways of addressing the excitation of well known maser emission from the 1720 MHz 
$F = 2^+ - 1^-$ hyperfine 
structure (hfs) satellite line in the $^2\Pi_{3/2}, J=3/2$ ground state of hydroxyl (OH)  in supernova 
remnants (SNRs) [\cite{4}] were presented [\ra G{\scriptsize RAY}]. Closely related, a careful theoretical 
and multi-line observational 
(absorption) study of OH excitation predicts maser emission in the 6049 MHz $J = 3^- - 2^+$ hfs satellite 
line of the rotationally excited $J=5/2$ state over a range of conditions thought to be prevalent in SNRs\footnote{Both the 1720 and the 6049 
MHz line connect, within
their respective rotational state, the hfs level with the highest energy and that with the lowest energy.},
which was, however, not detected despite a sensitive search. The results 
appear in conflict with the OH column density distributions predicted for some chemical models of SNRs 
[\ra W{\scriptsize ARDLE}]. 

1720 MHz OH masers are important tracers of the sub-mG magnetic fields of the 
$10^4$~cm$^{-3}$ density interstellar medium they are excited in. Density- and also temperature-wise 
($T \sim 100$~K) this post shock 
gas  might also be expected to produce class I methanol masers \textit{if} the elevated methanol 
abundance needed to produce high enough gain for observable maser emission can be attained (see \S \ref
{methanol}). However, so far searches for 36 and 44 GHz CH$_3$OH maser emission toward SNRs have had little 
success with the Sgr A East SNR near the Galactic center being a spectacular exception [\cite{5}]. Observations of 
a 
Fermi satellite $\gamma$-ray selected sample of SNRs only produced a couple of detections in these lines [\ra S{\scriptsize JOUWERMAN}].
Very interestingly, vice versa, narrow spectral features in the 1720 MHz OH line have been 
found at the velocities of more than 50 narrow class I methanol maser features in a survey of more than a hundred sources containing such masers; these are \textit{not} SNRs [\ra V{\scriptsize AL'TTS}].

\section{\label{bill}Bill Watson R. I. P.}
With great sadness, the maser community has observed the passing of Bill Watson, one of its most 
eminent theorists. Prof. William D. Watson, born January 12, 1942 in Memphis, TN, died on October 12, 
2009 in Urbana, where he had been doing research and teaching since 1972 at the University of Illinois. 
In the early seventies, after an important contribution on the formation of molecules on dust grain 
surfaces [\cite{6}], Professor Watson independently from and contemporaneously with another team  [\cite{7}],
established a chemistry driven by molecular ions as the basic paradigm of the gas phase chemistry of the 
interstellar medium [\cite{8}]. Ion-molecule chemistry explained the observed abundances of then newly detected 
molecules [\cite{9}] and predicted the observed strong deuterium enhancement as one of its natural consequences [\cite{10}]. After 
working on several other topics related to 
interstellar chemistry (e.g., [\cite{11}]), Bill Watson went on to become a world expert on astronomical masers and, 
particularly, on the radiative transfer of maser radiation and its polarization properties (e.g., [\cite{12, 13, 14}]). Together with a series of students, he completed a 
large number of studies on this very complex problem, significantly expanding fundamental work done in 
the early years of astronomical maser research. His work was always in unison with the forefront of 
observations. As an example, he correctly explained the origin of the high velocity features in the 
archetypical nuclear AGN maser NGC 4258 as originating in a $\sim 0.1$~pc sized ring, which was \textit
{subsequently} confirmed by Very Long Baseline Interferometry [\cite{15}].

\section{\label{bfield}Polarization and Magnetic Fields}
Models of molecular clouds in which magnetic fields are dynamically important (via ambipolar 
diffusion) predict that the relation between the magnetic field, $B$, and the density, $n$, is 
$B \propto n^\kappa$ with $\kappa \simless 0.5$. Observations (of the Zeeman effect) show that this relation holds over several orders of magnitude for $n > 100$
\ccm, below which $B$ is observed and predicted to be independent of $n$ and has a value of $\approx 
5~\mu$G [\cite{16}].
In particular, it holds for 1720 MHz OH masers ($n \sim 10^4$~\ccm, $B\sim0.2$~mG [\cite{17}]; see also \S 
\ref{theory}) and for the elevated densities required for  OH maser emission  in high mass star forming regions (HMSFRs)  ($n \sim 10^{6-7}$~\ccm, $B\sim$ a few to 
10~mG [\cite{18}]) and even  for 22.2 GHz H$_2$O masers ($n \sim 10^{8-9}$~\ccm, $B\sim$ tens of mG  [\cite{19}]). 
Masers are good $B$-field probes, even though most lines have a
small Zeeman splitting coefficient, i.e. the proportionality factor between the value of $B$ and the 
frequency (or equivalent velocity) shift between right and left circularly polarized signal. Under 
the extreme conditions of maser regions (high $n$ implies high $B$), $B$ fields are relatively easy 
to measure; the narrow line widths help. It is 
important to keep in mind 
that in many cases, whether a line is masing or not is a very sensitive function of density. For example, 
for many OH maser lines just a factor of two increase in density over the value at which optimal maser action 
occurs removes the inversion and, if a radio continuum background is present, absorption may be 
observed [\cite{20}]. In other words, within a certain range, Zeeman observations of all masers requiring a certain 
density for operation should deliver the same magnetic field strength. In particular, the values reported in 
the recent 
past for class I and class II methanol masers were orders of magnitude higher than expected, given the 
densities needed for their inversion (see \S \ref{methanol}) and fail to obey this requirement. Talks at the 
meeting make poorly known or unknown splitting coefficients responsible for this [\ra V{\scriptsize LEMMINGS}, 
S{\scriptsize ARMA}]. Theoretical calculations and laboratory measurements are urgently needed. 

An increasing amount of Very Long Baseline Interferometry (VLBI) results map $B$-field configurations in the vicinity of, mostly, high mass 
young stellar 
objects (YSOs) [\ra V{\scriptsize LEMMINGS}, S{\scriptsize URCIS}, C{\scriptsize HIBUEZE}], but now also of a
solar-like, low mass YSO.   The latter (VLA) observations, of H$_2$O maser emission in 
the 
famous IRAS 16293$-$2422, show that  the protostellar evolution of this object appears to be
magnetically dominated [\ra {\scriptsize DE}  O{\scriptsize LIVEIRA} A{\scriptsize LVES}].

Interesting polarization results for evolved stars include the finding that a dynamically significant and ordered $B$-
field is maintained over the whole of a circumstellar envelope [\ra A{\scriptsize MIRI}].  VLBI imaging of poloidal 
magnetic field morphologies at the launching site  of very high velocity water maser emission in so-called 
``water fountain'' sources indicate a magnetic origin of the observed bipolarity. These results could have 
far-reaching consequences on the long-standing question whether magnetic fields play a role in the 
symmetry breaking mechanism that transforms a circularly symmetric envelope into an axially symmetric 
planetary nebula.  [\ra A{\scriptsize MIRI}, I{\scriptsize MAI}, also C{\scriptsize LAUSSEN}, S{\scriptsize UAREZ}].

The question which of two different models that had been proposed to interpret circular polarization of SiO masers for the applicable case of weak Zeeman splitting is the correct one can be addressed by observations of  different maser transitions [\ra R{\scriptsize ICHTER}]. 

\section{Star Formation Masers}
The star formation session was dominated by presentations on methanol masers, which were the topic of 
a total of ten presentations in this and other sessions and ten posters. Below, I therefore dedicate some extra space to this molecule.

Other results on star forming regions include multi-epoch H$_2$O maser VLBI of two different deeply 
embedded objects in Cepheus A HW, tracing the motions of a bipolar outflow in one [\ra C{\scriptsize HIBUEZE}]
and, in the other, revealing what had appeared to be a spherical expanding shell, but seems to be 
showing more complex filamentary structure and  dynamics down to few AU scales [\ra T{\scriptsize 
ORRELLES}].

\subsection{Spinning the Big Spin -- Masers in Disks}
Masers in accretion disks have a long and checkered history. Consider, for example, one determines, 
via Gaussian fits, the centroid positions of two spatially well-separated maser 
spots, $A$ and $B$, peaking at distinctly different velocities, $v_m$ and $v_n$, in the velocity channels $m$ and $n$ of the correlator that have been observed (e.g., with 
the VLA) with an angular resolution that is much coarser than the separation of the spots. Then fits to the 
emission in velocity channels $i$, with $m < i < n$, will yield centroid positions lying between the real 
positions of the spots that seem to move monotonically from $A$ to $B$, mimicking a linear structure 
with a linear velocity gradient suggesting solid body rotation. This mis-interpretation of data gave rise to
 the \textit{Saga of the Rotating Methanol Maser Disks} [\cite{21,22}]\footnote{It remains a mystery why none of the disk 
 saga's proponents noticed that, in order to show 
solid body (and \textit{not} Keplerian) rotation the mass of the protostar would have to be 
negligible compared to the mask of the disk.}. Of course, such 
``disks'' are easily debunked by VLBI, which resolves the two maser spots into separate entities. 
Infrared $K$-band searches for shock-excited H$_2$ emission associated with the outflows 
expected in a disk scenario indeed found H$_2$ emission. However, in the great majority of the
cases, the H$_2$ emission was found to be elongated \textit{parallel} to the putative disks' orientation and not 
perpendicular to it as one would expect if there were a causal connection [\cite{23}].

VLBI indeed delivered the exciting picture of a 
rotating disk structure whose material giving rise to SiO maser emission appears also to be flowing in 
polar direction. The whole scenario, documented by a multi-Very Long Baseline Array-epoch movie, is that of an equatorially 
expanding ``excretion'' 
disk surrounding the peculiar radio source I in the Orion-KL region [\cite{24}], whose inner part is photo-ionized [\ra
G{\scriptsize REENHILL}]. 

Other bona fide maser emission from a disk, this time a completely photo-evaporating one, is that seen in (sub)millimeter 
and far-infrared hydrogen recombination line maser/ laser emission [\cite{25}] around the peculiar emission line B[e]star MWC 349 A, 
which excites a bipolar radio nebula [\cite{26}]. Fifteen year long monitoring of this object's polarization 
characteristics have yielded information on its magnetic field, while modeling its velocity 
distribution reveals Keplerian rotation in the disk's flaring outer parts and more complicated kinematics
in its inner regions  [\ra T{\scriptsize HUM}, B{\scriptsize  {\' A}EZ RUBIO}]. 

For a very long time, the MWC 349 A (sub)mm/far infrared recombination line 
maser/laser was one of its kind. It comes as a great relief that a second such source has now been found in 
Mon R2 IRS 2 [\ra J{\scriptsize IMENEZ}-S{\scriptsize ERRA}].

\subsection{\label{methanol}The Class I -- Class II Methanol Maser Dichotomy}
For more than 25 years it has being realized that strong interstellar methanol masers  come in two 
varieties [\cite{27,28}]: Class I methanol masers  arise from outflows from high mass protostellar objects 
(HMPOs) often significantly removed from the driving source, whereas Class II methanol masers (cIIMMs) 
arise from the nearest vicinity of the HMPO; sometimes forming conspicuous rings revealed by VLBI 
imaging [\ra B{\scriptsize ARTKIEWICZ} presented by {\scriptsize VAN} L{\scriptsize ANGEVELDE}].

To build up the gain for observable maser 
flux, a common 
requirement for both maser varieties is a substantial CH$_3$OH abundance. For cIIMMs masers this is 
achieved by the heating of dust grains in the dense HMPO envelope that evaporates methanol-containing ice  mantles and increases the gas phase methanol abundance by several orders of magnitudes. The 
warm grains, which attain an equilibrium temperature around 100--200 K emit intense mid/far infrared 
(IR) radiation that pumps the maser via torsional excitation\footnote{Note that the Planck function, $B_\nu(T)$, for 
$T = 200$ ~K, has its maximum at 11.75 THz, corresponding to a wavelength of 25.5~$\mu$m, very close 
to that of the main cIIMM pumping transition identified by modelers [\cite{29}].}. Interestingly, the dust emission from these 
so-called \textit{hot cores} is difficult or impossible to detect at near- or mid-IR wavelengths even when 
superior astrometry is achieved [\ra {\scriptsize DE} B{\scriptsize UIZER}]. Submillimeter continuum emission is 
frequently detected, but often with too coarse resolution to establish a clear association with the masers. 
Here, the shortest wavelength band of the Photodetector Array Camera and Spectrometer (PACS) on 
Herschel at $70~\mu$m promises to deliver crucial information [\ra P{\scriptsize ESTALOZZI}] on the maser host 
sources' spectral energy distribution and luminosity. 

Combining cIIMM proper motions from VLBI with 
interferometric observations of non-maser \meth\ (from the torsional ground and first excited state ($\sim 
400$~K above ground) will yield, in addition to $n$ and $T$, precious three dimensional velocity 
information on the closest vicinity of HMPOs [\ra T{\scriptsize ORSTENSSON}, {\scriptsize DE LA} F{\scriptsize UENTE}], 
allowing searches for infall, which so far is clearly indicated in one 
object (AFGL 5142) from VLBI observations alone [\ra G{\scriptsize ODDI}].

In complete opposition to cIIMMs, class I methanol masers (cIMMs) work in the \textit{absence} of a strong 
IR field, which means at significant offsets from HMPOs. An association of these masers with protostellar 
outflows has been established a long time ago [\cite{30}] and has recently been persuasively 
illustrated for very many sources 
by their coincidence with so-called extended green objects (EGOs), which are shocked regions
 [\ra {C{\scriptsize YGANOWSKI}]\footnote{EGOs or ``green fuzzies'' are regions of enhanced emission in images 
made with the InfraRed Array Camera (IRAC) on the Spitzer Space Observatory using the camera's 
$4.5~
\mu$m filter. Imaged in the course of the Galactic Legacy Infrared Mid-Plane Survey (GLIMPSE), the 
emission in this band is dominated by highly (shock-)excited lines of molecular hydrogen. The name 
derives from the fact that this emission was chosen to be color coded as green in false color 
presentations of multi-IRAC-band images [\cite{31}].}.

Also for a very long time, it has been known that the inversion of the observed cIMM lines  follows naturally,  over a range of physical conditions, from an 
interplay of both $E$- and $A$-type methanol's arrangement of energy levels in $k$-ladders and certain 
levels' transition probabilities (Einstein A-values) [\cite{32}]. This is confirmed by statistical equilibrium/radiative transfer calculations.  
%quite independent of (reasonably) assumed 
%collisional rate coefficients and recently confirmed by actually calculated values. 
Whether some cIMM lines show maser action in some regions and not in others depends (for temperatures of up to a few 
hundreds K) on the region's density, which has to be between $10^4$ and $10^5$ cm$^{-3}$, significantly lower 
than required for cIIMMs. Actually, under conditions conducive for cIMM action, but also in dark clouds, the strongest cIIMM lines, 
the 12.2 GHz $2_{0}-3_{-1}~E$ and 6.7 GHz $5_1-6_0~A^+$ transitions, are predicted to show enhanced 
absorption, which is actually observed in some regions [\cite{28,33}]\footnote{This enhanced absorption 
(``over-cooling'') follows from the fact that for $E$-($A$)-type methanol \textit{all} levels in the $k=-1$ 
($K=0$) energy ladder are overpopulated relative to levels in the neighboring $k=0$ ($K=1$) ladder. The 
same gives rise to the 
prominent $4_{-1}-3_0~E$ and $5_{-1}-4_0~E$ ($7_0-6_1~A^+$ and $8_0-7_1~A^+$) \textit{maser} 
lines near 36 and 84 GHz (44 and 95 GHz).}.

The physical conditions in the post shock gas hosting cIIMMs, $n$ and $T$ (for $B$ see below), can actually be quite well 
constrained just by the fact which of the 25 known cIMM lines (including a new one [\ra V{\scriptsize 
ORONKOV}, W{\scriptsize ALSH}, B{\scriptsize ROGAN}]) are masing and which are not. Here, extensive new surveys in the most prominent  transitions, at 44 and 25 GHz, deliver an abundance of new data  [\ra K{\scriptsize URTZ},  
B{\scriptsize YUN}, B{\scriptsize RITTON}]. In particular, the  
H$_2$O southern Galactic Plane Survey (HOPS), which apart from the 22.2 GHz  H$_2$O maser 
transition, covers several cI and cIIMM lines (plus multiple thermally excited lines from NH$_3$ and other 
species [\ra W{\scriptsize ALSH}]) has been most successful. Also, such surveys finally found the first long-
sought after 
methanol masers in \textit{low mass} star forming regions: Maser emission from cIMM lines was found in high
\meth\ abundance ``bullets'' of well known outflows driven by low mass protostars [\ra K{\scriptsize ALENSKII}]. Very interestingly, toward one of these, NGC 1333, the strongest cIIMM line 
($5_1-6_0~A^+$/6.7 GHz), has previously been found in absorption against the cosmic microwave background radiation [\cite{34}].

The elevated CH$_3$OH 
abundances required to produce an observable cIMM signal may also result, as for cIIMMs, from grain 
ice mantle desorption. However, in this case the necessary energy would be provided by a shock wave 
\textit{and not} by central heating from the HMPO. A high CH$_3$OH abundance  might even result from endothermic gas 
phase reactions, which require high temperatures (of order 10000 K), which may be reached in shocks [\cite{35}]. 
If cIMMs  arise in shock fronts, why do they virtually never show high velocity emission and have velocity spreads of just a few km/s? In contrast, 
H$_2$O masers, which are unequivocally associated with shocked outflows 
%via, first, their 
have much larger LSR velocity spreads of many tens, even up to hundreds of km/s.
%, and, second, their proper motions 
%determined with VLBI. Why, in contrast, do cIMMs only have velocity spreads of just a few km/s? 
The answer is likely that the cIMMs do barely have high enough methanol column densities to produce observable 
signal. Therefore, the geometry of a swept up shell seen edge on, i.e. from a direction perpendicular to 
that of the outflow motion would produce the longest coherent gain path in the direction of the observer. The radial velocity of emission from such a 
configuration would \textit{naturally} only have a 
small offset from the systemic velocity of the outflow source since the bulk of the outflow motion is in the plane of the sky. This scenario also implies that cIMM spots, 
partaking in the outflow, have large transverse motions, which for H$_2$O masers can be measured with 
VLBI. Unfortunately, cIMMs maser spots have all been found to be too large for successful VLBI [\cite{36}]. However, future proper motion measurements with the Karl G. Jansky Very Large Array (JVLA) may directly prove the above picture.

In this context it is noteworthy that magnetic field strength determinations of cIMMS via the Zeeman effect 
will provide measurements in an ISM  density regime barely covered by existing data and moreover 
deliver supremely important input for magnetohydrodynamic modeling of interstellar shocks. 
Unfortunately, as reported above (\S \ref{bfield}),  the 
$B$-field values reported so far (also at this meeting) are marred by our ignorance of the Zeeman 
splitting factors.     

\subsection{Periodic Methanol Masers}
The periodic variability of 6.7 and 12.2 GHz cIIMMs in a number of sources was firmly established by 
observations with South 
Africa's own Hartebeesthoek Radio Observatory by  our conference organizer Sharmila Goedhart in her 
dissertation [\cite{37,38}]. With periods from 20 to $>500$ days, this is one of the most peculiar phenomena in 
all maser science [\ra G{\scriptsize OEDHART}]. VLBI 
appears to rule out a periodic infrared radiation pump [\cite{39}], leaving a periodically varying 6--12 GHz radio 
continuum  background as a possibility. A model for this has been worked out in the framework of a 
colliding wind binary scenario providing ionizing radiation [\ra {\scriptsize VAN DEN} H{\scriptsize EEVER}, {\scriptsize VAN DER} W{\scriptsize ALT}].
\textit{If} a variable continuum background were at the heart of  cIIMM variability, this would raise the 
question whether \textit{all} cIIMMs need continuum photons as seeds to operate. While the first detections 
of these masers found the most prominent ones associated with ultracompact HII regions [\cite28}], in contrast, 
subsequent interferometric imaging surveys actually found that most cIIMMs had no associated 
continuum emission at the few mJy level (at 8.4 GHz) [\cite{40}]. Future, much more sensitive JVLA continuum 
surveys will address this question.

Finally, a comment on the chronology of masers in star forming regions, another perennial ``hen and 
egg'' topic of maser and star formation conferences [\ra B{\scriptsize REEN}]. Clearly, H$_2$O and cIMMs are found in outflows and 
thus accretion powered. CIMMs are frequently located quite far away from the outflow source (travel 
times of several ten thousands of years). Here we have the caveat that the transverse velocities are very 
uncertain and, given above arguments, may be quite a bit larger than the radial velocity spread, resulting 
in shorter time scales. In contrast, H$_2$O masers arise from within a few thousand AU (travel times a 
few hundred yr). The expansion time of an ultracompact HII region is thousands of yr, that of a 
hypercompact HII region hundreds of yr, comparable to the H$_2$O maser time scale. Critical 
questions are as follows: Before a HMPO has formed (started fusion): Can a H$_2$O maser outflow be driven,  
possibly by magnetically assisted disk-to-outflow angular momentum conversion? And, can accretion luminosity be sufficient to power a hot core and 
its  associated cIIMMs? Unequivocal evidence for 
disks and, in particular, high resolution imaging of outflow launching sites will address the first question, 
and evidence for the presence of radio emission at cIIMM positions the second.

\section{Stellar Masers}
The molecules producing circumstellar maser emission around oxygen-rich mass-losing evolved stars are either formed in or near the stellar photosphere (SiO,  H$_2$O) or in the expanding envelope (OH) by photo dissociation in its outer parts [\cite{41}]. These masers are observed, preferably, with interferometers [\ra R{\scriptsize ICHARDS}]. In addition, single dish studies give complementary interesting results, e.g., on the physical conditions 
in the extended atmosphere (by high excitation SiO maser line monitoring [\ra R{\scriptsize AMSTEDT}]) or on the 
magnetic field via polarization measurements; see \S\ref{bfield}. For  OH lines, such measurements, of over a hundred AGB and post-AGB stars, imply  $B = 0.2$--2.3 mG [\ra W{\scriptsize OLAK}]. Monitoring the phase lag of OH maser variability between signals arriving from the blue- and redshifted parts of the circumstellar shell, combined with angular shell size diameters from interferometry will deliver distances for $\sim 
20$ objects [\ra E{\scriptsize NGELS}]. 

IR interferometry and radio wavelength VLBI have enjoyed great synergy 
when IR imaging found the molecular regions just outside the photospheres of oxygen-rich Asymptotic 
Giant Branch stars (AGBs) (so-called MOLspheres [\cite{42}]) [\ra W{\scriptsize ITTKOWSKI}]  at the same 
distance from the stellar surface as SiO masers. In many cases, the SiO masers form beautiful rings, 
strongly 
indicating tangential amplification [\ra C{\scriptsize OTTON}, A{\scriptsize L MUTAFKI}, D{\scriptsize ESMURS}]. SiO 
masers have the potential of probing the magnetic fields, via polarization measurements [\ra C{\scriptsize 
OTTON}], and also the dynamics in these interesting regions, as demonstrated by the spectacular movie 
made from 112 epochs of SiO maser Very Long Baseline Array (VLBA) imaging [\ra G{\scriptsize ONIDAKIS}].

\section{\label{mega}Cosmology and the Hubble Constant: AGN and Megamasers}

After a concise, but comprehensive introduction to the \textit{Standard Model of Cosmology}, a status report on 
the 
Megamaser Cosmology Project (MCP) was given [\ra H{\scriptsize ENKEL}]. The goal is to measure the Hubble 
constant, $H_0$, with a precision of a few percent  to complement the existing cosmic microwave 
background radiation data in placing constraints on the nature of Dark Energy.  The MCP's targets are active 
galaxies with 22.2 GHz  H$_2$O maser mission originating from a $< 1$ pc region around the central 
supermassive Black Hole. Best suited are systems seen nearly edge on, such as the ``Golden 
Source'' NGC 4258 (see \S \ref{bill}). Comparing the maser distributions'  rotational speeds and radii, determined from 
VLBA imaging, with the measured centripetal acceleration (from spectral monitoring) directly delivers 
the systems' distances, $D$,  which can be compared with the measured  redshifts, yielding $H_0$ and also 
the Black Hole mass [\ra W{\scriptsize ARDLE}]. Naturally, systems with $D > \sim 50$ Mpc are desirable, which 
are well partaking in the Hubble flow (and whose redshifts are not significantly influenced by local 
dynamics). Recently, a beautiful new example of such as system was found, UGC 3789, for which a 
distance of 50 Mpc was determined and an even farther system, NGC 6262 at 
$D = 152$ Mpc. Together, these two yield a Hubble constant of $67 \pm 6$~km~s$^{-1}$Mpc$^{-1}$.
[\ra H{\scriptsize ENKEL}]. Another nice system is Mrk 1419 [\ra I{\scriptsize MPELLIZZERI}]. 

For NGC 4258 itself, we 
were shown what sophisticated modeling of 18 epochs of VLBI observations and 10 years of single dish 
monitoring can do for you: a detailed analysis of disk warping and elliptical orbits of 
maser clumps with differential precession [\ra H{\scriptsize UMPHREYS}] and, on top of all this, a highly precise 
distance. 

To date, several thousand galaxies have been surveyed for  H$_2$O megamasers, yielding just $
\approx 
130$ detections. Obviously, increasing the sample is  highly desirable as is any criterion that could help 
to increase 
the success rate. Cross-matching galaxies with  H$_2$O megamaser detections with systems found 
in the  Sloan Digital Sky Survey yields an increased maser detection rate for galaxies showing strong 
[OIII] $\lambda 5700$ emission. [\ra Z{\scriptsize AW}].

Apart from nuclear disks,  H$_2$O megamasers have been found in outflows from AGN and in the jets' 
interaction zones with the interstellar medium and new detections were reported. [\ra T{\scriptsize ARCHI}].

Interesting progress was also reported for  H$_2$O ``kilomasers'', which are found in star burst galaxies. 
Spectacular JVLA imaging revealed such masers in several active locations in the merging Antennae system (NGC 4038/4039), likely marking the birth sites of super star clusters. In particular, its 80 times higher  bandwidth and hugely increased number of spectral channels (compared to the VLA) makes the JVLA  a tremendously efficient survey 
machine for extragalactic maser emission [\ra D{\scriptsize ARLING}]. 

OH megamasers in the central starbursts 
of Ultra Luminous Infra Red Galaxies  probably mark the most extreme star forming conditions in the local 
Universe. These masers even ''contaminate'' blind surveys for 21 cm emission from HI. Finding such 
systems and also formaldehyde (H$_2$CO) megamasers at higher redshift  is highly desirable, among others in a 
H$_2$CO Deep Field [D{\scriptsize ARLING}, B{\scriptsize AAN}], but have so far been unsuccessful (for OH at $z>1$) 
[\ra W{\scriptsize ILLETT}]. Here the Five hundred meter Aperture Spherical Telescope, currently being built in 
south western China, holds great promise for the foreseeable future [\ra J. Z{\scriptsize HANG}]. FAST will have almost three times the collecting area of the Arecibo 300 m telescope, the existing OH megamaser detection machine, and a \textit{much} larger sky coverage [\cite{43}].

\section{Maser Astrometry}
Already at the previous maser conference a whole series of contributions reported high precision 
multi-epoch VLBI astrometry (mostly) of masers in HMSFRs, yielding accurate distances 
and proper motions. In the meanwhile this field has greatly expanded and about 50 parallaxes obtained 
with the Japanese Very Long Baseline Exploration of Radio Astronomy Array (VERA)\footnote{http://
veraserver.mtk.nao.ac.jp/outline/index-e.html} and the Bar and Spiral Structure Legacy survey 
(BeSSeL)\footnote{http://www.mpifr-bonn.mpg.de/staff/abrunthaler/BeSSeL/index.shtml} 
using the VLBA have established the location of Outer Galaxy spiral arms and even resulted in a revision 
of the Galactic rotation parameters [\cite{44}] [\ra R{\scriptsize EID}, H{\scriptsize ONMA}, S{\scriptsize AKAI}, M{\scriptsize ATSUMOTO}]. So far, 
mostly 22.2 GHz  H$_2$O masers and  12.2 GHz cIIMMs  [\ra X{\scriptsize U}] have been employed. Astrometry with the much stronger 6.7 GHz cIIMMs has started with the European VLBI Network (EVN) and VERA and will soon (in mid-2012) be possible  with the VLBA. 

Sources discussed at our meeting include well and (up to 
now) not so well studied HMSFRs, for example  ON 1 and 2 [\ra N{\scriptsize AGAYAMA}], W33 [\ra I{\scriptsize MMER}]. and  the 
prominent W51Main/South  region, for which H$_2$O and 6.7 GHz cIIMM astrometry allows a comparison of the sources from which emission in the different species arises, outflows (with measured expansion motions) traced by H$_2$O  [\ra S{\scriptsize ATO}]  and hot cores by \meth\  [\ra E{\scriptsize TOKA}].  

In addition, astrometry for the famous protoplanetary ``Rotten Egg'' 
nebula OH $231.8$ $+4.2$ [\ra C{\scriptsize HOI}] and other post AGB stars [\ra I{\scriptsize MAI}] as well as the classical OH/IR hypergiant NML Cyg was reported [\ra B. Z{\scriptsize HANG}].
So far, SiO masers have not yet played a major role in VLBI/Galactic structure astrometry efforts. However, in the future this may change 
a great deal thanks to the extensive surveys for vibrationally exited (\textit{v} = 1 and 2), $J=1-0$ masers 
conducted with the Nobeyama 45 meter telescope that have led to the detection of well over 1000 
sources [\ra D{\scriptsize EGUCHI}]. These masers, around 43 GHz, can be observed with VERA and the VLBA.

Other surveys will find many more methanol and water masers 
(see \S\ref{methanol}). In particular, an interferometric follow-up of the 6.0/6.7 GHz Parkes/ATCA/MERLIN multi-beam survey in the 22.2 GHz H$_2$O line will certainly detect many new water masers associated with high mass 
star formation in the general vicinity of the methanol masers. However, it also has the potential of 
detecting 
H$_2$O masers associated with low mass YSOs have formed together with the high mass YSOs, 
probably in clusters. In fact, the luminosity of known such masers can be high enough to make them detectable at distances of many kpc and make them the \textit{only} signposts for low mass star formation 
outside of the Solar neighborhood [\ra T{\scriptsize ITMARSH}].

\section{Odds and Ends: Other Masers, Propagation/Scattering, New Facilities}
\subsection{Formaldehyde and Ammonia Masers}
Maser action from molecules other than OH, H$_2$O and SiO has been discovered, at centimeter wavelengths, in many lines from (mostly) non-metastable levels
of ammonia (NH$_3$) and from formaldehyde (H$_2$CO). 
NH$_3$ and H$_2$CO masers have been exclusively found 
in the hot cores around HMPOs, in the close vicinity, but generally not coincident with H$_2$O masers and cIIMMs 
[\cite{45,46}] [\ra M{\scriptsize 
ENTEN}, B{\scriptsize ROGAN}]. 
Interestingly, the H$_2$CO maser line is the 4.8 GHz $1_{10} - 1_{11}$ $K$-doublet transition, which is 
ubiquitously found almost always in absorption throughout our Galaxy [\cite{47}] (and others [\cite{48}]), even in the diffuse ISM [\cite{49}] . While the anti-inversion (over-cooling) leading to enhanced absorption is well understood [\cite{50}], the excitation of  maser emission in this line is not [\cite{51}].  It strikes 
me as peculiar that  in  all of the $\sim10$ known maser sources the emission is very weak: The flux densities of most sources are around 0.1 Jy or smaller. The most 
luminous one known, in Sgr B2, ($\approx 0.5$~Jy), has a luminosity that is roughly 100 times lower than that of the strongest 6.7 GHz cIIMM in that region [\cite{52,53}].

%\subsection{Ammonia Masers}
Maser emission in the $(J,K)  = (3,3)$ inversion line of ortho NH$_3$\footnote{Ortho-NH$_3$ assumes states, $J,K$, with $K = 0$ or $3n$, where $n$ is an integer (all H spins 
parallel), whereas $K \ne 3$ for para-NH$_3$  (not all H spins parallel). 
The principal quantum numbers $J$ and $K$ correspond to the total angular momentum and its projection on the symmetry axis of the pyramidal molecule.} has been 
known for a long time [\cite{54}]. In very few outflow sources [NGC 6334 I and DR21(OH)] it has been shown that this, and sometimes also the (6,6)  line, share properties of 
cIMM lines found in the same region, i.e., identical location of maser spots and a narrow, single component profile [\cite{55,56}]. In fact, this is the \textit{only} known molecular line 
emission with a one-to-one correspondence to cIMM emission. In contrast, numerous (mostly weak) maser lines have been found from \textit{non-metastable} ammonia levels [\cite{57}], 
even in its rare $^{15}$NH$_3$ isotopopologue, which is more than 200 times less abundant than $^{14}$NH$_3$
[\cite{58,59}].  Non-metastable levels, with $J>K$, decay rapidly 
down 
their $K$-ladders until they arrive at the lowest levels (with $J=K$). These are metastable and form a thermal distribution, which is the reason for ammonia's fame as a 
molecular cloud thermometer. Given the high transition probabilities of the FIR rotational lines connecting them, the non-metastable levels' populations are strongly influenced by the mid 
IR continuum, which, first, gives rise to maser action in certain lines and, second, places,  as for cIIMMs, the maser's emission regions close to HMPOs. 

\subsection{Submillimeter Hydrogen Cyanide and Water Masers}
In the (sub)millimeter range,  HCN lines from within several vibrationally excited states have been found to be masing (or lasing) 
[\cite{60,61,62}] (plus the $J=1 - 0$ line from the ground state [\cite{63}]). To these we may 
add H$_2$O masers in the vibrational ground state and the excited bending
 mode [\cite{64,65,66}]. The emission region of the vibrationally excited lines is pretty clear: The 
exceedingly high energies above the ground state (up to more than 4000 K places 
their origin very close to the stellar photospheres of mass 
losing stars with carbon-rich (HCN) and oxygen-rich (H$_2$O) chemistry, respectively.

Until a very short time ago, the only means to get information about the astrophysically very important 
water molecule was observing \textit{maser} emission in lines emitted from high energies above the 
ground, which are are not sufficiently excited in the Earth's atmosphere to make it completely opaque at 
and around their frequencies. Earlier space missions either observed, with modest angular resolution, 
only the H$_2$O $1_{10} - 1_{01}$ ground-state line  near 557 GHz (\textit{SWAS} [\cite{67}] and \textit{Odin} [\cite{68}]) or far- and mid-
IR  lines with limited angular and spectral resolution (ISO; e.g., [\cite{69}]). This situation has dramatically changed for the better 
with \textit{Herschel}, which 
produces a wealth of H$_2$O data, observing lines between 500 and 1400 GHz with excellent sensitivity and spectral 
resolution [\cite{70}]. However, one should keep in mind that interferometric observations of the maser lines 
accessible from the ground are the \textit{only} means to get information on this molecule with an 
angular resolution better than $10''$  [\ra M{\scriptsize 
ENTEN}].

\subsection{ Maser Propagation and New Facilities}
Observations of short time scale maser variability caused in part by (even anisotropic) interstellar scattering may provide information on the intervening interstellar medium, but also on the 
intrinsic line of sight dimensions of individual maser spots [\ra L{\scriptsize ASKAR}, D{\scriptsize ESHPANDE}, Mc{\scriptsize 
CALLUM}]. 
 
Exploring the nature of all masers will greatly benefit from the comprehensive or greatly expanded collecting area and/or frequency coverage and much larger number of spectral channels available with the Atacama Large Millimeter 
Array (ALMA)\footnote{http://www.almaobservatory.org/}  and the 
JVLA [\ra B{\scriptsize ROGAN}, W{\scriptsize OOTTEN}], e-MERLIN\footnote{http://www.e-merlin.ac.uk/} and 
the Australia Telescope Compact Array (ATCA)\footnote{http://www.narrabri.atnf.csiro.au/}, 
with MeerKAT\footnote{http://www.ska.ac.za/meerkat/} [\cite{71}] playing a role as well [\ra B{\scriptsize OOTH}].

\section{Famous Last Words}
Our field is like the (unsaturated) maser process -- it stimulates itself and grows very rapidly! A few remarks:

\noindent
\begin{itemize}
\item Big surveys are  going on, but results need to be digested! Maser surveys need to be cross 
correlated with radio, IR and dust continuum surveys. In order to be useful, e.g., for planning VLBI observations, positions determined need to be listed with realistic uncertainty estimates.
\item Think big! Compared to their predecessors, the new and upgraded facilities, JVLA, ALMA, e-MERLIN, and ATCA, offer awesome advances in \textit{much} wider band correlator capability. Therefore, when planning your observation, make sure to use all the capabilities at your disposal. For, example, observe not just ``your'' target maser line, but cover as much bandwidth as possible, e.g., to get 
good continuum sensitivity or to cover other interesting lines. This is called ``commensal'' (= symbiotic) observing and hopefully will be a policy adopted and encouraged by observatories. Disk space is cheap and gets ever cheaper!
\item Much more astrometry is needed! VERA and VLBA/BeSSeL run at full tilt! The EVN can also do astrometry. 
What about the Japanese VLBI and the East Asian VLBI Networks? We need VLBA-like capability in the southern hemisphere!
\item If possible, all maser VLBI observations should use phase referencing, even if astrometry is not the goal. The resulting \textit{absolute} position information is indispensable for comparative studies.
\item It is worth to emphasize projects that have an impact on \textit{all} of astronomy and even beyond, namely those that address Galactic structure, Local Group dynamics,
precision $H_0$, protostellar collapse, magnetic fields: cIMMs could probe MHD shocks; the launching of protostellar outflows/shaping of planetary nebulae and other phenomena.
\end{itemize} 

Finally, as a veteran of all the meetings in the series, with all my heart I thank the organizers of this one very much for a perfect conference and the most pleasant time I had attending it. 

I'm thankful to Christian Henkel and Mark Reid for their comments on the manuscript.

{}

\end{document}